\documentclass{elsart}

\usepackage{amssymb}
\usepackage{graphicx}

\newcommand{\RR}{{\bf R}}

\newcommand{\kk}{\mathbf{k}}
\newcommand{\rr}{{\bf r}}

\newcommand{\BE}{\begin{equation}}
\newcommand{\EE}{\end{equation}}
\newcommand{\BEN}{\begin{eqnarray}}
\newcommand{\EEN}{\end{eqnarray}}

\newcommand{\LL}{{\mathbf L}}

\newcommand{\beq}{\begin{equation}}
\newcommand{\eeq}{\end{equation}}
\newcommand{\bea}{\begin{eqnarray}}
\newcommand{\eea}{\end{eqnarray}}
\newcommand{\fwidth}{3.5in}

\begin{document}

\begin{frontmatter}



{\scriptsize \sf \vspace*{-40pt}Proceedings article of the Study of Matter at Extreme Conditions (SMEC) conference in Miami, Florida\\ submitted to {\em Journal of Physics and Chemistry of Solids} (2005)\\[10pt]}

\title{Simulations of Dense Atomic Hydrogen in the Wigner Crystal Phase}

 

\author{Burkhard Militzer}
\address{Geophysical Laboratory, Carnegie Institution of Washington, 5251 Broad Branch Road, N.W., Washington, DC 20015}
\author{Rebekah L. Graham}
\address{College of Creative Studies, University of California, Santa Barbara, CA 93106}

\begin{abstract}
Path integral Monte Carlo simulations are applied to study dense
atomic hydrogen in the regime where the protons form a Wigner
crystal. The interaction of the protons with the degenerate electron
gas is modeled by Thomas-Fermi screening, which leads to a Yukawa
potential for the proton-proton interaction. A numerical technique for
the derivation of the corresponding action of the paths is
described. For a fixed density of $r_s=200$, the melting is analyzed
using the Lindemann ratio, the structure factor and free energy
calculations. Anharmonic effects in the crystal vibrations are
analyzed.
\end{abstract}
\begin{keyword}

Path integral Monte Carlo \sep  hydrogen melting \sep Wigner crystal\sep pair action


\PACS 
\end{keyword}
\end{frontmatter}

\section{Introduction}

Recent advances in high pressure experiments and computer simulations
techniques have led to substantial progress in our understanding of
the melting properties of solid hydrogen at high pressure. Using
diamond anvil cell experiments, E. Gregoryanz {\em et
al.}~\cite{Gregoryanz03} extended the experimental determination of
the melting line from 15 GPa~\cite{Datchi00} to 40 GPa. Bonev {\em et
al.} combined the two-phase melting technique with {\em ab initio}
simulations~\cite{Ogitsu04} and predicted that there is a maximum in
the melting temperature around 80 GPa~\cite{BonevNature}. At these
pressures, hydrogen is still in molecular form. The question of
interest is whether the melting temperature keeps decreasing as the
pressure is increased further, or if another phase appears and the
melting temperature again increases.

From theoretical arguments, we know that atomic hydrogen eventually
transforms into a Wigner crystal where the protons form a body
centered cubic (b.c.c.) lattice. The focus of this work is to
characterize this regime more accurately, to determine its melting
line and to analyze by how much the known molecular solids phase and
the Wigner crystal phase are separated on the pressure or density
scale.

In this article, we study the Wigner crystal of protons using path
integral Monte Carlo (PIMC). This technique allows us to characterize
the quantum effects of the protons. The anharmonic effects in the
lattice vibrations are also included accurately. While at much higher
temperature and lower density, one can describe the electrons from
first principles~\cite{MC00}. For this work, we instead approximate
the electron-proton interaction using Thomas-Fermi theory. This leads
to an effective Yukawa potential for the proton-proton interaction,
\beq
V_Y(r) = \frac{\mathbf{e}^2}{r} e^{-r/D_s}\;,
\eeq
where the screening length $D_s$ is given by~\cite{AshcroftMermin},
\beq
D_s = \left( \frac{\pi}{12} \right)^{1/3} \left(\frac{r_s}{a_0}\right)^{1/2} \, a_0 \;,
\eeq 
and $r_s$ is the Wigner-Seitz radius, $\frac{4}{3}\pi r_s^3 =
V/N$. Throughout this work, we will use units of {\em nuclear} Bohr
radii, $a_0 = 4\pi\epsilon_0\hbar^2 / m_p \mathbf{e}^2$ =
2.9$\times$$10^{-14}$~m and {\em nuclear} Hartrees, Ha = $\mathbf{e}^2 /
(4\pi\epsilon_0a_{0})$ = 8.0$\times$$10^{-15}$~J = $5.8\times
10^{8}$~K $k_b$, which are by a factor by $m_p/m_e=1836$ shorter, or
larger respectively, than the usual atomic units.

Jones and Ceperley~\cite{JC96} used PIMC to study quantum melting in
Coulomb systems where the electrons were assumed to form a rigid
background. We extend their work by introducing the Yukawa potential
in order to understand how the electronic screening affects the
stability of the Wigner crystal.

\section{Path integral Monte Carlo}

The thermodynamic properties of a many-body quantum system at finite
temperature can be computed by averaging over the density matrix,
$\hat{\rho} = e^{-\beta \hat{H}}, \beta=1/k_{\rm b} T$.  
Path integral formalism~\cite{Fe53} is based on the identity,
\beq
\hat{\rho} \equiv e^{-\beta \hat{H}} = \left[ e^{-\frac{\beta}{M} \hat{H}} \right]^M \;,
\eeq
where $M$ is a positive integer. Insertion of complete sets of states
between the $M$ factors leads to the usual imaginary time path
integral formulation, written here in real space,
\beq
\rho(\RR,\RR';\beta)=
\int\ldots\int d\RR_{1}\ldots d\RR_{M-1} \; \rho(\RR,\RR_{1};\tau)\ldots\rho(\RR_{M-1},\RR';\tau) \;,
     \label{eq2.5}
\eeq
where $\tau=\beta/M$ is the time step, and $\RR$ is a collective
coordinate including all particles, $\RR=\{\rr_1,\ldots,\rr_N\}$. Each
of the $M$ steps in the path now has a high temperature density matrix
$\rho(\RR_k,\RR_{k+1};\tau)$ associated with it. The integrals are
evaluated by Monte Carlo methods. For the densities under
consideration, we can neglect exchange effects of the protons and
represent them by distinguishable particles. Given these constraints,
PIMC is an exact technique and free of uncontrolled approximations
(assuming the Yukawa potential is valid). This technique includes the
correct phonon excitations in the presence of anharmonic effects,
which we will discuss later.

\subsection{Action for an Isolated Pair of Particles}

The action plays a central role in PIMC since it determines the
weights of paths. We will describe a novel approach for its
derivation, which we found to be more accurate for Yukawa systems than
previous techniques. First we discuss the action for an isolated pair
of particles, and then we introduce periodic boundary conditions
commonly used in many-body simulations.

Typically, one approximates the high-temperature many-body density
matrix,\\ $\rho(\RR,\RR';\tau)$, as a product of exact pair density
matrices which can be motivated using the Feynman-Kac (FK) formula,
\bea
\frac{\rho(\RR,\RR';\tau)}{\rho_0(\RR,\RR';\tau)} 
&=& 
\left< e^{-\int_0^\tau dt \sum_{i<j} V_Y(\rr_{ij}) } \right>_{\rm \RR \to \RR'}
=
\left< \prod_{i<j} e^{-\int_0^\tau dt V_Y(\rr_{ij}) } \right>_{\rm \RR \to \RR'}\\
&\approx&
\prod_{i<j} \left< e^{-\int_0^\tau dt V_Y(\rr_{ij}) } \right>_{\rm \rr_{ij} \to \rr_{ij}'} 
\equiv
e^{ - \sum_{i<j} U_Y(\rr_{ij},\rr_{ij}'; \tau) },
\eea
where $\rho_0(\RR,\RR';\tau)$ is the free particle density matrix.
$U_Y(\rr_{ij},\rr_{ij}'; \tau)$ is the pair action corresponding to
all paths separated by $\rr_{ij}$ at imaginary time $t=0$ and by
$\rr'_{ij}$ at $t=\tau$. An approximation is introduced when one makes
the assumption that the different pair interactions can be averaged by
independent Brownian random walks, denoted by brackets
$\left<\ldots\right>$. The pair action approximation is exact for two
particle problem. However, higher-order correlations are left out,
which must be recovered in the many-body PIMC simulations using a
sufficiently small time step $\tau$.

The pair action, $U_Y$, can be computed by three different methods. 1)
For certain potentials where the eigenstates are known in analytical
form, e.g. for the Coulomb potential, the action can be derived from
the sum of state~\cite{Po88}. However to our knowledge, for the Yukawa
potential they are not known analytically. 2) In the matrix squaring
technique~\cite{St68}, one represents the density matrix on a grid and
successively lowers the temperature by performing a one-step path
integration. This method can be applied to arbitrary potentials. 3)
For the Yukawa potential, we found it advantageous to use the FK
formula to derive the pair action. Computationally, it is a bit more
expansive than matrix squaring but it does not introduce grid errors
that we found difficult to control in case of the Yukawa
potential. The FK approach is also applicable to arbitrary potentials
unless they exhibit an attractive singularity, which is discussed
further in~\cite{MP05}.

In FK approach one derives the action $U_Y(\rr,\rr';\beta)$
stochastically by generating an ensemble of random paths according to
the free particle action $U_0(\rr,\rr';\beta)$ that begin at $\rr$ and
terminate at $\rr'$,
\beq
e^{-(U_Y-U_0)} = \left< e^{-\tau/2 \sum_{i=1}^M [V_Y(\rr_{i-1}) + V_Y(\rr_{i})] }\right>_{U_0(\rr,\rr';\beta)}
\label{FK} \;.
\eeq
The free-particle paths can be generated by a bisection
algorithm~\cite{Ce95}. We found that $M=\beta/\tau=32$ time steps was
sufficient for the Yukawa potential.

To determine the kinetic energy in simulations, one also needs the
derivative of the action with respect to $\beta$, which can be
evaluated from the same set of paths,
\begin{eqnarray}
\frac{\partial}{\partial \beta}(U_Y-U_0) 
=
\left< e^{ - \sum_i \tau V_Y(\rr_i)} 
\frac{1}{M}\sum_{i=1}^{M} \left[ V_Y(\rr_i) + \frac{1}{2}(\rr_i-\rr_i^{\rm Cl}) \cdot \nabla V_Y(\rr_i) \right]
\right>_{U_0(\rr,\rr';\beta)}.
\end{eqnarray}
where $\rr_i^{\rm Cl}$ represents the classical path between the two end
points.

\begin{figure}[htb]
\centerline{\includegraphics[angle=0,width=\fwidth]{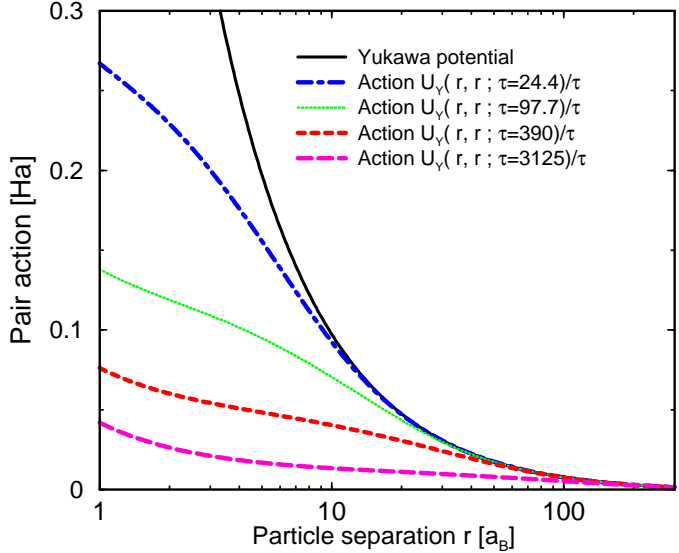}}
\caption{The pair action $U_Y(r,r'=r;\tau)$ scaled by by $1/\tau$
       is compared with the corresponding Yukawa potential as a
       function path separation, $r$. The Yukawa screening length of
       $D_s=387$ represents the electronic screening of the proton
       interaction in hydrogen at a density of $r_s=200$. The action
       converges to the primitive approximation, Eq.~\ref{prim}, for
       large $r$ and small $\tau$. At small $r$, quantum fluctuations
       remove the singularity present in the Yukawa potential and lead
       to a linear dependence on $r$, which is known as the cusp
       condition.}
\label{action}
\end{figure}

The FK formula yields the action for only one specific pair of $\rr$
to $\rr'$. For the diagonal action, $\rr=\rr'$, we map out a whole
grid of points beginning from a small value near the origin to large
values (several times the thermal de Broglie wavelength given by
$\sqrt{2 \pi \hbar^2 \beta/m}$). Typically, we use a logarithmic grid
with about 500 points. For large $r$, the action approaches the
classical limit given by the primitive approximation,
\beq
U_Y(\rr,\rr';\beta) \approx
\frac{\beta}{2}\left[ V_Y(\rr) + V_Y(\rr') \right]\;,
\label{prim}
\eeq
which is shown in Fig.~\ref{action}.

Statistical uncertainties in the resulting action are intrinsic to the
FK approach. We found that $10^6$ paths yield sufficiently small error
bars. However, any noise in tabulated action values is impractical for
the subsequent interpolation in PIMC simulations. To eliminate this
problem, we use the same random paths for all grid points in the
table. This does not remove the uncertainty but prevents noise in the
tables.

Including off-diagonal density matrix elements in PIMC simulations
allows one to use larger time steps, which makes the simulation more
efficient. However, the off-diagonal terms are more difficult to
obtain with the FK approach, which is one of the limitations of the
approach. Here, we only consider the leading term in an
expansion of the action~\cite{Ce95},
\beq
U_Y(\rr,\rr';\beta) \approx U_Y(q,q;\beta) + s^2 \: \xi(q;\beta) \; + \; \ldots
\eeq 
where $q=\frac{1}{2}(|\rr|+|\rr'|)$ and $s=|\rr-\rr'|$. It turns out
that it is more efficient to derive $\xi$ from finite differences in
$s$ rather than evaluating an analytical expression. Having completed
our derivation for the action of an isolated pair of particles, we now
consider a system with periodic boundary conditions.

\subsection{Pair Action in Periodic Boundary Conditions}

In our PIMC simulations, we use $N=250$ particles, which are initially
placed on the sites of a b.c.c. lattice. For the density under
consideration ($r_s=200$), the corresponding screening length for
hydrogen ($D_s=387.667$) is comparable in magnitude to the length of
the simulation cell $L=1218.59$. Consequently, long-range effects from
periodic image particles are far from negligible, and significant care
must be taken to derive results in the thermodynamic limit.

The total potential energy for a system of $N$ particles
interacting via the Yukawa potential $V_Y(r)$ is given by~\cite{AT87},
\beq
V = \sum_{i>j} \sum_\LL V_Y(\rr_{ij}+\LL)
\; + \;
\frac{1}{2} \sum_i \sum_{\LL \ne 0} V_Y(\LL)  \;,
\label{pot2}
\eeq
where $\rr_{ij}=\rr_i-\rr_j$ and $\LL$ is a lattice vector. Instead of
representing long-range terms, $V_P(\rr) = \sum_\LL V_Y(\rr+\LL)$, on
a 3D table~\cite{hamaguchi}, we adopt the optimized Ewald~\cite{Ew17},
technique by Natoli and Ceperley~\cite{Na95} and express the potential
as a sum of one real-space image, $W(|\rr|)$, and
a number of Fourier components,
\beq
V_P(\rr) \equiv \sum_\LL V_Y(\rr+\LL) \approx  W(|\rr|) + \sum_{|\kk|\leq k_c} y_\kk e^{+i\kk\rr}.
\label{breakup}
\eeq
Following Natoli and Ceperley~\cite{Na95}, we express $W$ as a
linear combination of fifth-order polynomials, $W(|\rr|)=\sum_n
a_n f_n(|\rr|)$, which is known as a locally piecewise-quintic Hermite
interpolant. The fit coefficients $a_n$ and $y_\kk$ can be derived by
minimizing the $\chi^2$ deviation,
\beq
\chi^2 = \frac{1}{\Omega} \int_\Omega d^3\rr \left[ V_P(\rr) - \sum_n a_n f_n(|\rr|) - \sum_{|\kk|\leq k_c} y_\kk e^{+i\kk\rr} \right]^2 \;,
\label{chi2}
\eeq
where $\Omega=L^3$ is the volume of the unit cell. For the Fourier
coefficients this directly yields,
\beq
y_\kk = v_\kk - \sum_n a_n f_{n\kk}\;, ~~~~~~\mbox{with}~~~~~~ v_\kk = \frac{1}{\Omega} \int_\Omega d^3\rr V_P(\rr) e^{-i\kk\rr} \;,
\eeq 
where $v_\kk$ and $f_{n\kk}$ are the corresponding Fourier
transforms. Deviating from~\cite{Na95}, we derive the coefficients
$a_n$ from the following set of linear equations, $m=\{1,\ldots,n\}$,
\beq
\left[ v_m - \sum_{|\kk|\leq k_c} v_\kk f_{m\kk} \right] 
  = \sum_n a_n \left[ f_{nm} - \sum_{|\kk|\leq k_c} f_{n\kk} f_{m\kk} \right].
\eeq
$v_m$ and $f_{nm}$ are overlap integrals, $f_{nm} = \frac{1}{\Omega} \int_\Omega d^3\rr f_n(\rr) f_m(\rr)$, and
\bea
v_m &=& \frac{1}{\Omega} \int_\Omega d^3\rr \; V_P(\rr) \, f_m(|\rr|) = \frac{1}{\Omega} \sum_\LL \int_\Omega d^3\rr \; V_Y(\rr) \, f_m(|\rr|)\\
&=& \frac{4 \pi}{\Omega}  \int dr \, r^2 \, V_Y(r) f_m(r) \; + \; \sum_{\LL \ne 0} \frac{2 \pi}{\Omega L} \int dr \, r \, f_m(r) \int_{L-r}^{L+r} dq \, q \, V_Y(q) \;.
\label{real}
\eea
In the last expression, one must sum over a sufficiently large number
of images until the interaction is completely screened, $|\LL| \gg
D_s$. Computing the coefficients $a_n$ using the real-space
integration in Eq.~\ref{real} is more efficient and accurate than the
Fourier integration employed in~\cite{Na95}. Our approach also works
well for the Coulomb problem, which was the motivation for
the~\cite{Na95} work. In this case, the Ewald potential replaces $V_P$
in Eq.~\ref{chi2}.

This optimized Ewald approach provides us with an accurate
representation of the periodic functions leading to efficient
many-body simulations. We apply it to the Yukawa potential, $V_Y$, to
the corresponding pair action, $U_Y$, and also to the kinetic energy
term, $\left(\frac{d U_Y}{d \beta} - V_Y \right)$, unless it is very
small for $r=L/2$. Typically, we use between 10 and 20 shells of $k$
vectors.

\section{Results}

\begin{figure}[htb]
\centerline{\includegraphics[angle=0,width=\fwidth]{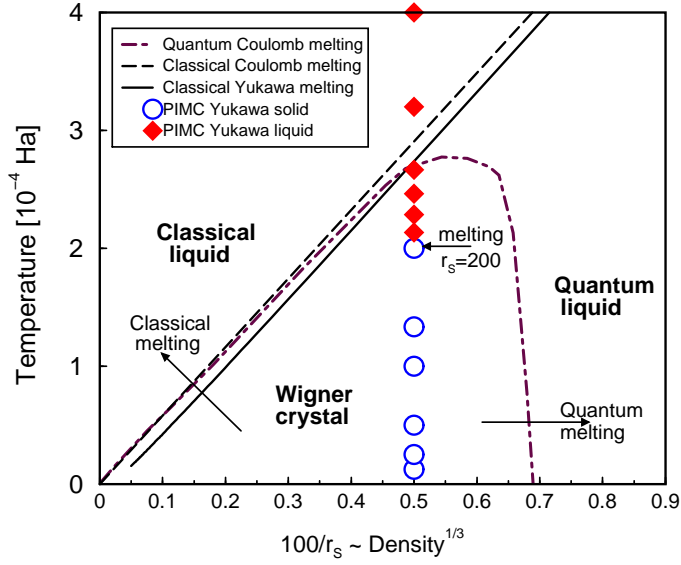}}
\caption{Phase diagram for Wigner crystal. The dashed 
	lines show the classical melting line, $\Gamma=172$, and
	quantum melting line for Coulomb systems computed by Jones and
	Ceperley~\cite{JC96} with PIMC. The solid lines show the
	classical melting line for the Yukawa system with the
	screening length chosen for hydrogen. Our PIMC simulations at
	$r_s=200$ ($\circ$ and $\diamond$) show a significant
	reduction in the melting temperature below the classical value
	due to the quantum effects of the protons.}
\label{phase}
\end{figure}

The phase diagram in Fig.~\ref{phase} relates our simulations at a
fixed density of $r_s=200$ to the classical Yukawa melting computed by
Hamaguchi, the classical Coulomb melting line given by
$\Gamma=\frac{\mathbf{e^2}}{r_sk_bT}=172$, and the quantum melting for
Coulomb systems~\cite{JC96}. 

\begin{figure}[htb]
\centerline{\includegraphics[angle=0,width=\fwidth]{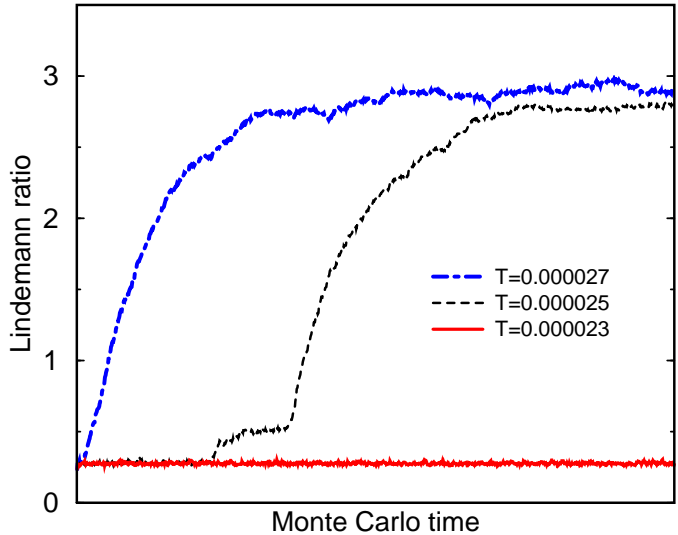}}
\caption{The evolution of the Lindemann ratio 
        is shown for the three Monte Carlo simulations in the vicinity
        of the melting temperature. }
\label{linde}
\end{figure}

The most straightforward way to detect melting in the simulation is to
monitor the instantaneous value of the Lindemann ratio, $\gamma =
\sqrt{\left<u^2\right>}/r_{NN}$, which relates the average
displacement of a particle from its original lattice site to the
nearest neighbor distance. Fig.~\ref{linde} shows the Lindemann ratio
for three Monte Carlo simulations. At the beginning of each 
simulation, the particles are in the classical b.c.c. ground
state. For temperatures sufficiently above the melting line, the
system melts instantly. For temperatures only slightly above the
melting line, the simulation shows a meta-stable superheated solid,
which might melt at some point during the MC simulation, as the black
dashed line indicates. The time it takes to melt depends not only on
temperature, but also on system size, the type of MC moves, and the MC
random numbers, thereby making this criterion impractical for determining
the melting temperature.

\begin{figure}[htb]
\centerline{\includegraphics[angle=0,width=\fwidth]{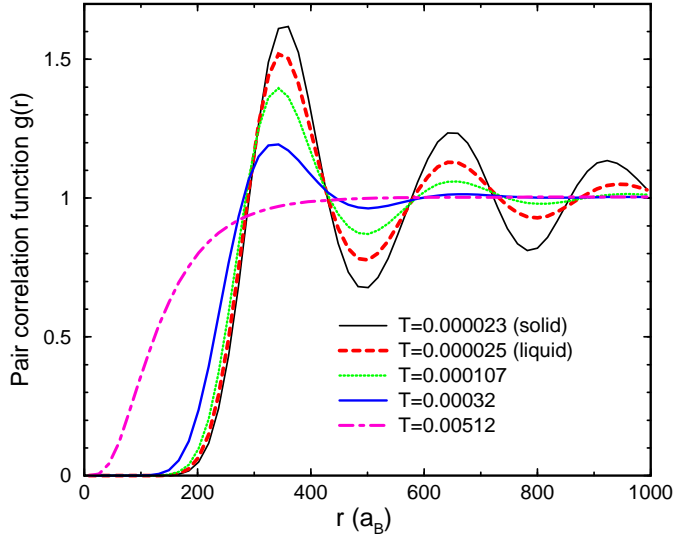}}
\caption{Pair correlation functions, $g(r)$, for different temperatures. }
\label{gr}
\end{figure}

Fig.~\ref{gr} shows a series of pair correlation functions, $g(r)$,
for simulations at different temperatures. The magnitude of the
oscillations in the $g(r)$ show a significant temperature dependence
in the liquid phase. However, there is only a small change upon
melting and all $g(r)$ functions in the solid phase are practically
identical to the example shown for T=2.3$\times$10$^{-5}$.

\begin{figure}[htb]
\centerline{\includegraphics[angle=0,width=\fwidth]{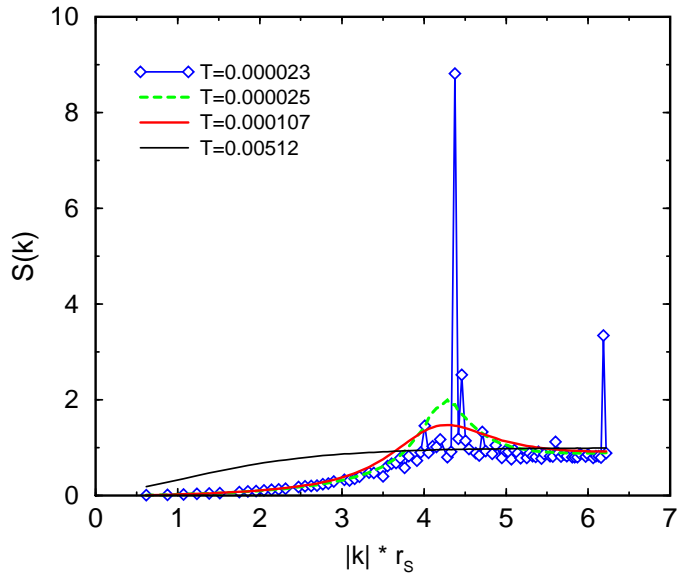}}
\caption{The sharp peaks in the structure factor, $S(k)$, disappear when 
        the system melts and a pattern typical for a liquid
        appears. This transition coincides with the increase in the
        Lindemann ratio.}
\label{sk}
\end{figure}

Compared to the pair correlation function, the structure factor,
$S(k)$, shows significant changes upon melting (Fig.~\ref{sk}). The
disappearance of the peaks coincides with the increase in the
Lindemann ratio beyond stable value of approximately 0.28. However,
neither method can determine whether a simulation is in the
meta-stable state. They only lead to an upper bound of the melting
temperature. To determine the thermodynamic phase boundary, one needs
the free energy in both phases, which can be obtained through
thermodynamic integration of the internal energies.

\begin{figure}[htb]
\centerline{\includegraphics[angle=0,width=\fwidth]{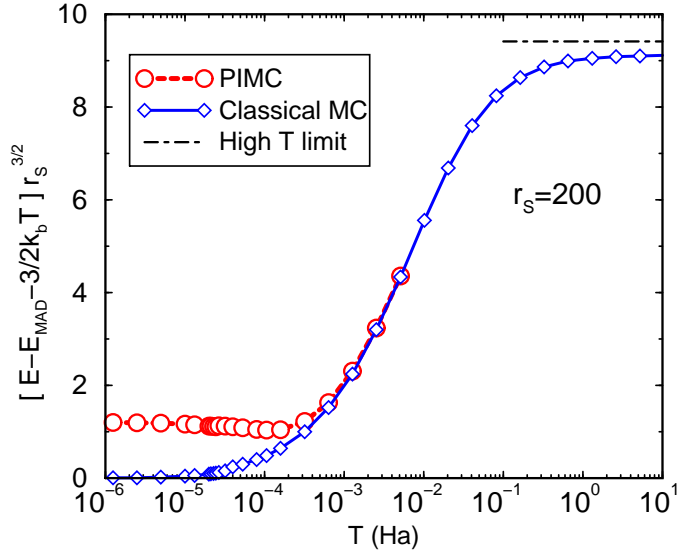}}
\caption{The internal energy per particle is shown as a function of 
       temperature. The classical kinetic energy and the Madelung term
       have been removed. }
\label{E1}
\end{figure}

Fig.~\ref{E1} compares PIMC internal energies with results from
corresponding classical MC that we have performed. At high temperature
when the thermal de Broglie wavelength is short compared to the
inter-particle spacing, the protons behave classically and the PIMC
energies approach results from classical MC simulations. At low T,
both result differ substantially due to the zero point motion. The
analytical high T limit~\cite{hamaguchi} is not reached exactly, due
to finite size effects.

\begin{figure}[htb]
\centerline{\includegraphics[angle=0,width=\fwidth]{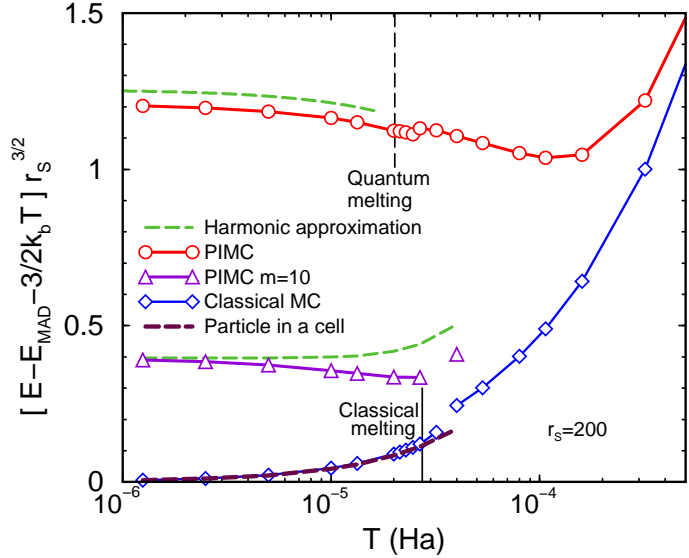}}
\caption{Low temperature region of Fig.~\ref{E1}. Results from the 
        harmonic lattice approximation and from a classical
        particle-in-a-cell model have been added. Furthermore, PIMC
        and harmonic results for particles with mass=10 instead of 1
        have been included to demonstrate that the harmonic
        approximation becomes more accurate as zero point fluctuations
        are reduced.}
\label{E2}
\end{figure}

In Fig.~\ref{E2}, we compare MC results with our results derived from
the harmonic lattice approximation and the particle-in-a-cell (PIC)
model. In the PIC approximation, one assumes that the thermal motion
of the particles are uncorrelated. One freezes all particles in the
supercell except one and derives all thermodynamic variables from the
motion of this single classical particle. The PIC internal energies
agree remarkably well with the corresponding classical MC results. A
generalization to the quantum case is not straightforward. If one
simply considers a quantum particle, represented by a path in a
lattice of frozen particles then the resulting kinetic energies are
far too high (worse than the harmonic approximation) because the
remaining classical particles provide too high of a confining force due
to the missing quantum fluctuations.

\begin{figure}[htb]
\centerline{\includegraphics[angle=0,width=\fwidth]{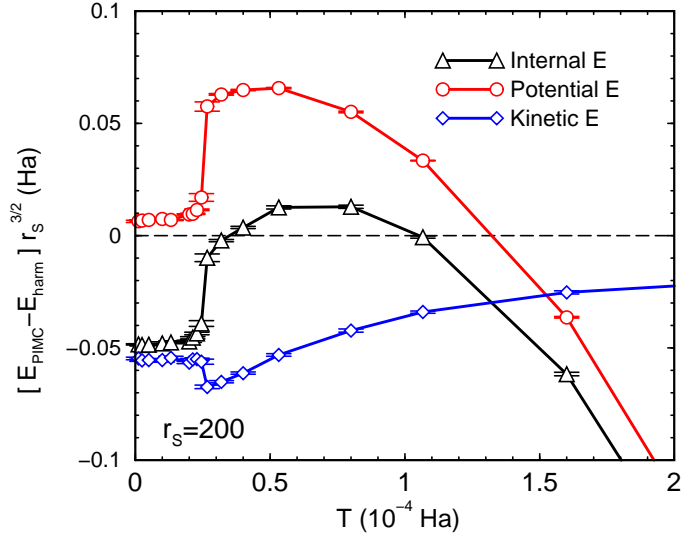}}
\caption{Difference in internal, kinetic, and potential energy between 
       PIMC and the harmonic lattice approximation. The sharp kinks
       near T=2.5$\times$10$^{-5}$ indicate the melting transition.}
\label{harm}
\end{figure}

Fig.~\ref{E2} also shows results from the harmonic lattice
approximation for a unit cell of corresponding size. Harmonic internal
energies are significantly overestimated, primarily due to errors in
the kinetic energy as demonstrated in Fig.~\ref{harm}. The zero point
motion of the protons is large enough so that paths travel into
regions of the potential where the harmonic approximation is not
longer valid. A comparison of the PIMC Lindemann ratios and the
harmonic values shows that the harmonic approximation localizes the
particles too much, thereby increasing the kinetic energy.

To further support this conclusion, we perform PIMC and harmonic
calculation with particles 10 times as heavy. Fig.~\ref{E2} shows
that the agreement improves substantially.

\begin{figure}[htb]
\centerline{\includegraphics[angle=0,width=\fwidth]{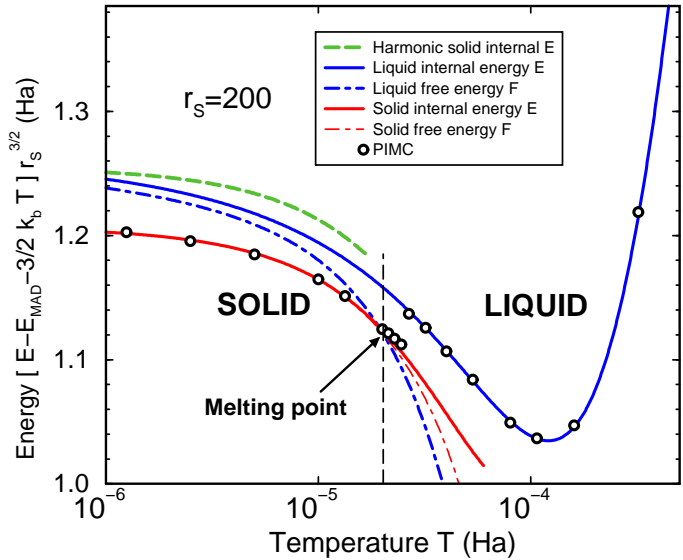}}
\caption{Internal and free energies, $E$ and $F$, are shown for the solid 
       and liquid phase.}
\label{E3}
\end{figure}

Fig.~\ref{E3} shows the internal PIMC energies along with the free
energies obtained from thermodynamic integration. The solid free
energies agree very well with solid internal energies until melting,
which suggest that each phonon mode is in its ground state, and
excitation in the phonon spectra leads directly to melting. At low T,
the free energies of the crystalline state are lower than the
extrapolated values for the liquid, which means the solid phase is
stable and the density of $r_s=200$ is not yet high enough to reach
the quantum melting transition (see Fig.~\ref{phase}).

At T=2.0$\times$10$^{-5}$ the free energies of both phases match,
which determines the thermodynamic melting point. This corresponds to
a temperature of 11,700 K and a hydrogen mass density of 2,100
g$\,$cm$^{-3}$.

Fig.~\ref{E3} also shows that a number of simulations that appeared to
be stable were actually meta-stable, which can also occur in classical
simulations (Fig.~\ref{E2}). The melting temperature of
2.0$\times$10$^{-5}$ is significantly below the corresponding
classical value of 2.7$\times$10$^{-5}$, which suggests that quantum
effects are important. However, a finite size extrapolation remains to
be done.

\section{Conclusions}

In this article, we used many-body computer simulations of protons
interacting via a Yukawa potential to model dense atomic hydrogen in
the regime of the Wigner crystal. Path integral Monte Carlo
simulations were employed to capture the quantum effects of the
protons. Electronic screening effects were treated in the Thomas-Fermi
approximation, which distinguishes our results from the earlier work
by Jones and Ceperley~\cite{JC96}.  

We use the Lindemann ratio, pair correlation functions, and the
structure factor to study the stability of the Wigner crystal and to
detect melting. We observed that the system can remain in a
meta-stable state of a super-heated crystal during the entire course
of a PIMC simulation, which makes a direct determination of the
melting temperature very difficult.

Instead, a reliable melting temperature can be obtained by matching of
the free energies of both phases, which were derived by thermodynamic
integration of the PIMC internal energies. For the density under
consideration, $r_s=200$, we found that the quantum Yukawa systems
melts at significantly lower temperatures than the corresponding
classical system.

Furthermore, we compared our PIMC results with other more approximate
techniques. The harmonic lattice approximation overestimates the
kinetic energies significantly because the zero point motion of the
protons is strong enough so that anharmonic effects in the crystal
field become relevant. We also compared with a classical
particle-in-a-cell model and found good agreement with classical MC
simulation. However this method cannot be generalized easily to the
case of quantum protons.

We plan to extend our analysis to other densities and to derive a
phase diagram that indicates the stability of the Wigner crystal of
nuclei in the presence of electronic screening effects. A careful
analysis of finite size effects also remains to be done. Future
theoretical work on dense atomic hydrogen will need to describe the
electronic properties on a more fundamental level. Coupled
ion-electron Monte Carlo is one promising approach~\cite{PCH04}.

\section{Acknowledgments}

We would like to acknowledge stimulating discussions with J. Kohanoff
and M. Magnitskaya. R.L.G. received support from the National Science
Foundation's Research Experiences for Undergraduates program at the
Carnegie Institution of Washington.




\begin{thebibliography}{10}

\bibitem{Gregoryanz03}
E.~Gregoryanz, A.~F. Goncharov, K.~Matsuishi, H.~Mao, and R.~J. Hemley.
\newblock {\em Phys. Rev. Lett.}, 90:175701, 2003.

\bibitem{Datchi00}
F.~Datchi, P.~Loubeyre, and R.~LeToullec.
\newblock {\em Phys. Rev. B}, 61:6535, 2000.

\bibitem{Ogitsu04}
Tadashi Ogitsu, Eric Schwegler, Francois Gygi, and Giulia Galli.
\newblock {\em Nature}, 91:175502, 2003.

\bibitem{BonevNature}
S.A. Bonev, E.~Schwegler, T.~Ogitsu, and G.~Galli.
\newblock {\em Nature}, 431:669, 2004.

\bibitem{MC00}
B.~Militzer and D.~M. Ceperley.
\newblock {\em Phys. Rev. Lett.}, 85:1890, 2000.

\bibitem{AshcroftMermin}
N.~W. Ashcroft and N.~D. Mermin.
\newblock {\em Solid State Physics}.
\newblock Harcourt, Inc., Orlando, FL, 1976.

\bibitem{JC96}
M.~D. Jones and D.~M. Ceperley.
\newblock {\em Phys. Rev. Lett.}, 76:4572, 1996.

\bibitem{Fe53}
R.~P. Feynman.
\newblock {\em Phys. Rev.}, 90:1116, 1953.

\bibitem{Po88}
E.~L. Pollock.
\newblock {\em Comp. Phys. Comm.}, {52 }:49, 1988.

\bibitem{St68}
R.~G. Storer.
\newblock {\em J. Math. Phys.}, {9}:964, 1968.

\bibitem{MP05}
B.~Militzer and E.~L. Pollock.
\newblock {\em Phys. Rev. B}, 71:134303, 2005.

\bibitem{Ce95}
D.~M. Ceperley.
\newblock {\em Rev. Mod. Phys.}, 67:279, 1995.

\bibitem{AT87}
M.P. Allen and D.J. Tildesley.
\newblock {\em Computer Simulation of Liquids}.
\newblock Oxford University Press, New York, 1987.

\bibitem{hamaguchi}
S.~Hamaguchi, R.~T. Farouki, and D.~H.~E. Dubin.
\newblock {\em Phys. Rev. E}, 56:4671, 1997.

\bibitem{Ew17}
P.P. Ewald.
\newblock {\em Ann. Phys.}, {54}:557, 1917.

\bibitem{Na95}
V.~Natoli and D.~M. Ceperley.
\newblock {\em J. Comp. Phys.}, 117:171--178, 1995.

\bibitem{PCH04}
C.~Pierleoni, D.~M. Ceperley, and M.~Holzmann.
\newblock {\em Phys. Rev. Lett.}, 93:146402, 2004.

\end{thebibliography}





\end{document}